\documentclass[aps,prl,twocolumn,groupedaddress]{revtex4}
\usepackage{graphicx}

\newcommand{\nnu}{$\nu_{tot}= 1$}

\newcommand{\Rpar}{$R^{\parallel}_{xx}$}
\newcommand{\RHpar}{$R^{\parallel}_{xy}$}
\newcommand{\RCF}{$R^{CF}_{xx}$}
\newcommand{\RHCF}{$R^{CF}_{xy}$}
\newcommand{\SCF}{$\sigma^{CF}_{xx}$}
\newcommand{\Spar}{$\sigma^{\parallel}_{xx}$}
\newcommand{\rhopar}{$\rho^{\parallel}_{xx}$}
\newcommand{\rhoCF}{$\rho^{CF}_{xx}$}

\begin{document}

\title{Vanishing Hall Resistance at High Magnetic Field in a Double Layer Two-Dimensional Electron System}

\author{M. Kellogg$^1$, J.~P. Eisenstein$^1$, L.~N. Pfeiffer$^2$, and K. W. West$^2$}

\affiliation{$^1$California Institute of Technology, Pasadena CA 91125 
\\
	 $^2$Bell Laboratories, Lucent Technologies, Murray Hill, NJ 
07974\\}

\date{\today}

\begin{abstract}At total Landau level filling factor \nnu\ a double layer two-dimensional electron system with small interlayer separation supports a collective state possessing spontaneous interlayer phase coherence.  This state exhibits the quantized Hall effect when equal electrical currents flow in parallel through the two layers.  In contrast, if the currents in the two layers are equal, but oppositely directed, both the longitudinal and Hall resistances of each layer vanish in the low temperature limit. This finding supports the prediction that the ground state at \nnu\ is an excitonic superfluid. 

\end{abstract}

\pacs{73.40.-c, 73.20.-r, 73.63.Hs}

\maketitle

The Hall effect is possibly the most widely observed phenomenon in solid state physics\cite{Hall}.  Owing to the Lorentz force, an ordinary current-carrying conductor in a magnetic field develops a voltage drop $V_H$ transverse to both the current $I$ and the field. This Hall voltage is directly proportional to the magnetic field and offers a useful measure of the concentration and sign of the constituent particles responsible for conduction in the material.  In more exotic systems the Hall resistance $R_H=V_H/I$ can deviate strongly from this simple dependence. For example, in two-dimensional electron systems (2DES) $R_H$ exhibits ranges of magnetic field over which it is constant and precisely equal to the quantum of resistance $h/e^2$ divided by either an integer\cite{klitzing} or certain rational fractions\cite{tsui}. These quantized Hall effects (QHE) reflect the presence of an energy gap in the system, due either to Landau quantization of the cyclotron orbits of the electrons or to the strong interactions between electrons in a partially filled Landau band. In superconductors, where a coherent collective electronic state is present, the Hall resistance vanishes altogether\cite{supercon}. 

In this paper we report low temperature measurements of the Hall and longitudinal resistances of a double layer 2DES in a strong perpendicular magnetic field $B$.  Our focus is on the situation in which the total electron density $N_{tot}$ of the double layer system is equal to the degeneracy $eB/h$ of a single spin-resolved Landau level produced by the magnetic field. It is well known\cite{perspectives} that in this $\nu_{tot} = hN_{tot}/eB = 1$ case the system possesses an unusual strongly correlated phase when the separation between the two layers is less than a critical value.  Owing to interlayer Coulomb interactions the critical layer separation remains non-zero even in the limit of arbitrarily weak tunneling between the layers.  In this zero tunneling limit the electron system exhibits a quantized Hall effect\cite{murphy} with $R_H = h/e^2$, a dramatically enhanced and sharply resonant zero bias tunneling conductance\cite{spielman1}, and exact quantization of the Hall component of Coulomb drag between the layers\cite{kellogg1}.  It is well-established theoretically that the many-electron state responsible for these phenomena possesses an unusual broken symmetry: spontaneous interlayer phase coherence.  This collective phase may be viewed in several equivalent ways, including as a Bose condensate of interlayer excitons\cite{macd1} or a pseudospin ferromagnet\cite{yang}. In addition to the fascinating phenomena already observed in this collective phase, there remains the prediction that the system should possess counterflow superfluidity: equal but oppositely directed currents in the two layers should produce no dissipation (in linear response) at low temperatures\cite{wenzee,moon,ezawa,stern}.  Dissipationless transport is, of course, a hallmark of the ordinary QHE.  However, in that case the longitudinal conductivity vanishes along with the resistivity owing to the large Hall resistance.  In contrast, in the \nnu\ interlayer coherent phase, the longitudinal conductivity in counterflow is expected to be infinite.  At high magnetic field this can only happen if both the longitudinal {\it and} Hall resistances in counterflow vanish.  It is this effect which we report here. 
 
\begin{figure}
\centering
\includegraphics[width=2.5 in,bb=217 628 395 727]{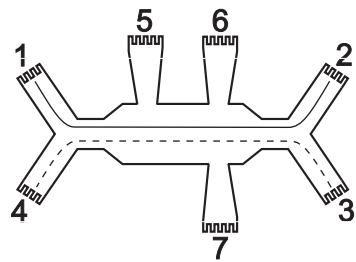}
\caption{\label{fig:fig1} Schematic drawing of mesa structure confining the 2DES.  Arms 1, 2, 3, and 4 are for injecting and withdrawing current, while arms 5, 6, and 7 are for measuring voltages.  Solid line indicates current pathway through one 2DES layer, dashed line the pathway in the other layer.  Gates are not shown.}
\end{figure}

The sample used in these experiments is a conventional GaAs/AlGaAs heterostructure. Two 18nm GaAs quantum wells are separated by a 10nm $\rm Al_{0.9}Ga_{0.1}As$ barrier layer.  Each well contains, in the sample's as-grown state, a 2DES with carrier density $N_1 \approx N_2 \approx 5.4\times10^{10}\rm cm^{-2}$ and a low-temperature mobility of about $\rm 1\times 10^6 cm^2/Vs$.  
The 2DES is confined to a bar-shaped mesa, shown schematically in Fig. 1, whose central region is $160{\rm \mu m}$ wide and $320{\rm \mu m}$ long. At each end of this bar the width narrows down to $80{\rm \mu m}$ before branching into two separate pathways.  The resulting four arms (labeled 1, 2, 3, 4 in the figure) are used to inject and withdraw current from the two 2DES layers in the sample.  Three voltage probe arms (labeled 5, 6, 7) extend away from the central bar and allow measurements of the Hall and longitudinal voltages.  Diffused AuNiGe contacts are placed at the ends of each of the arms.  Though not shown in the figure, gate electrodes on the front and back of the thinned sample cross these arms and enable separate electrical contact to the individual 2DES layers\cite{jpe1}.  Front and back gate electrodes, also not shown, cover the central bar and allow for independent control of the densities of the two 2DESs.  

Resistance measurements are performed using 0.5 nA, 2.3Hz excitation and standard lock-in detection. The excitation current is first injected into one layer and then withdrawn from it before being redirected into the second layer. The redirection may be done at room temperature where the choice between parallel and counterflow transport is made by selecting the appropriate mesa arm for injecting the current into the second layer. By comparing the current injected into the first layer with the amount available for redirection into the second, we can determine how much current tunnels from one layer to the other inside the sample.  For the data presented here this leakage never exceeds 1\% of the total transport current, even deep inside the \nnu\ interlayer coherent phase where interlayer tunneling is strongly enhanced\cite{spielman1}.  This experimental configuration assures that the magnitudes of the currents in the two layers are essentially identical in counterflow, even if leakage leaves them slightly less than the total current injected into the sample. Finally, we emphasize that the longitudinal and Hall voltage drops in the system are measured in one of the two layers, typically the top layer.  This is done to avoid creating current shunts between the layers at the location of the voltage probes. Although quantitative differences between the layer voltages are observed, they are small and do not alter any of the conclusions of this work.

\begin{figure}
\centering
\includegraphics[width=3 in,bb=152 327 381 513]{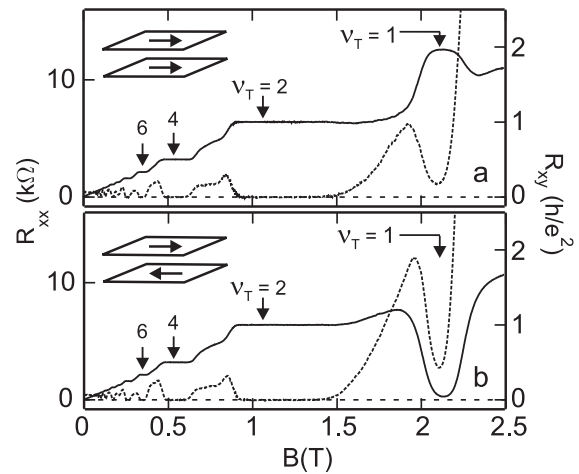}
\caption{\label{fig:fig2} Hall and longitudinal resistances (solid and dotted traces, respectively) in a low density double layer 2DES at $T$ = 50 mK. a) Currents in parallel in the two layers. b) Currents in counterflow configuration.  Resistances determined from voltage measurements on one of the layers.}
\end{figure}

Figure 2a shows the magnetic field dependence of the longitudinal and Hall resistances, \Rpar\ and \RHpar\ respectively, at $T$ = 50 mK with equal currents flowing in parallel through the two layers. These resistances are computed by dividing the appropriate voltages by the current flowing in the individual layers. For the data in the figure, the densities of the individual 2DESs have been reduced to $N_1 = N_2 = 2.54 \times 10^{10} \rm cm^{-2}$ using the front and back center gates.  At this density the ratio of the center-to-center separation of the quantum wells, $d$ = 28 nm, to the magnetic length $\ell = (\hbar/eB)^{1/2}$ at \nnu\ is $d/\ell = 1.58$.  This value is small enough that the double layer 2DES at \nnu\ should be well within the QHE phase\cite{kellogg2}. This is confirmed by the well-developed minimum in \Rpar\ and the flat plateau in \RHpar\ around \nnu\ at $B = 2.1 \rm T$ in the figure. Since the tunnel-induced splitting between the lowest symmetric and antisymmetric double well eigenstates in this sample is estimated to be only about $\Delta_{SAS} \sim 0.1$ mK, while the mean inter-electron Coulomb energies are roughly $10^6$ times larger, this \nnu\ QHE state should be well-approximated by the spontaneously interlayer phase coherent excitonic (or pseudo-ferromagnetic) model. 

Note that the \nnu\ Hall plateau in Fig. 2a occurs at \RHpar\ = $2h/e^2$. This is twice the value mentioned above simply because we define the resistance as the voltage divided by the current flowing in a single layer, not the net current flowing through the bilayer. In addition to this intrinsically bilayer QHE, numerous single layer QHE states, e.g. at $\nu_{tot}$ = 2, 4, 6, etc., are also evident in the data. 

Figure 2b illustrates our main result.  The data in this figure were taken under the same conditions as that in Fig. 2a, except that the currents in the two layers flow in opposite directions.  In this counterflow configuration much of the data appears very similar to that obtained in the parallel configuration.  For example, at low magnetic fields and around the single layer QHE states at $\nu_{tot}$ = 2, 4, 6, etc., the counterflow resistances \RCF\ and \RHCF\ are very similar to the parallel flow resistances \Rpar\ and \RHpar. At \nnu\ this similarity persists in the case of the longitudinal resistances \Rpar\ and \RCF; both exhibit a deep minimum near $B$ = 2.1T.  In contrast, however, the Hall resistances are dramatically different.  While \RHpar\ is quantized at $2h/e^2$, the counterflow Hall resistance \RHCF\ exhibits a deep local minimum.  We reiterate that this Hall resistance is measured with voltage probes connected to only one of the 2D layers in the system; the small value of \RHCF\ {\it does not} result from a cancellation of opposite sign Hall effects in two layers shorted together at the voltage contacts. 

\begin{figure}
\centering
\includegraphics[width=3.0 in,bb=187 77 414 241]{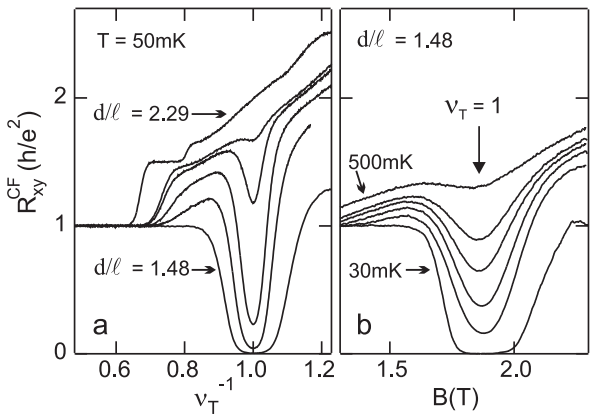}
\caption{\label{fig:fig3} Development of deep minimum in \RHCF\ with decreasing effective layer separation (panel a) and falling temperature (panel b). In a) data taken at various $d/\ell$ (1.48, 1.59, 1.66, 1.71, 1.75, and 2.29) are plotted versus inverse filling factor $\nu^{-1}_{tot}$. In b) the fixed $d/\ell$ data, taken at $T = 30$, 150, 200, 250, 300, and 500 mK, are plotted versus magnetic field.
}
\end{figure}

Figure 3a demonstrates that the minimum in the counterflow Hall resistance at \nnu\ develops rapidly as the effective layer separation $d/\ell$ is reduced below about $d/\ell \approx 1.8$.  This is not surprising since previous tunneling \cite{spielman1} and Coulomb drag \cite{kellogg1} measurements on samples taken from the same semiconductor wafer as the present one have established that the critical point separating the strongly-coupled excitonic \nnu\ QHE state from a weakly-coupled non-QHE phase occurs very close to this value. As $d/\ell$ is decreased further the minimum in \RHCF\ at \nnu\ deepens, falling to essentially zero by $d/\ell = 1.48$.  Figure 3b shows that this remarkable transport feature weakens as the temperature is increased, becoming only a shallow local minimum by $T$ = 500 mK.

Figure 4 summarizes our measurements of the temperature dependences of all four relevant resistances, \Rpar, \RHpar, \RCF, and \RHCF, at \nnu\ and $d/\ell = 1.48$. Figure 4a shows the measured temperature dependences of \Rpar\ and \RHpar\ from $T$ = 400 mK down to about 35 mK. Over this range the Hall resistance remains nearly constant at \RHpar\ = $2h/e^2$ while the longitudinal resistance vanishes in a thermally activated fashion: \Rpar\ $\sim R_0 e^{-\Delta/2T}$ with the energy gap $\Delta \approx 0.5 \rm K$.  Thus, the behavior of the parallel flow transport at \nnu\ is qualitatively the same as that of any ordinary QHE state. 

Figure 4b displays the temperature dependence of the counterflow resistances \RCF\ and \RHCF.  Both quantities appear to vanish in the low temperature limit. The two resistances are surprisingly similar in magnitude over most of the temperature range.  The general temperature dependence of each is less clearly thermally activated than \Rpar. \RHCF, in particular, shows significant curvature on the Arrhenius plot in the figure. We note in passing that quantitative variations in the various resistances were encountered.  We attribute these to the disorder in the sample which has been observed to change upon thermal cycling and repeated strong gating of the 2DES densities. Indeed, as Fig. 2 makes clear, the \nnu\ QHE state occurs amidst an otherwise rapid approach to an insulating state at high magnetic field.  This and other indications suggest that disorder is quite important in these samples.

The data described above vividly demonstrate that it is possible for both the longitudinal and Hall components of the resistivity tensor of a bilayer 2DES to vanish at high magnetic field when oppositely directed currents flow in the two layers.  This result is consistent with the expectation that the \nnu\ bilayer QHE state is an excitonic superfluid.  This unusual quantum fluid is believed to possess two distinct dissipationless transport mechanisms. First, in parallel transport current is carried through the sample by charged quasiparticle excitations lying near the edges of the sample. This mode of transport is dissipationless, but only in the conventional QHE sense: both the longitudinal resistance \Rpar\ {\it and the conductivity} \Spar\ vanish as $T \rightarrow 0$.  Second, the \nnu\ excitonic state is also expected to possesses a coherent transport mechanism within its condensate. This mechanism may be viewed as dissipationless transport of charge neutral excitons or, equivalently, counterflowing charge currents in the individual layers.  Not surprisingly, neutral excitons feel no Lorentz force and thus \RHCF\ is expected to vanish along with \RCF. The expectation is that the longitudinal conductivity in counterflow, \SCF, should be infinite.  

\begin{figure}
\centering
\includegraphics[width=3.25 in,bb=161 111 426 277]{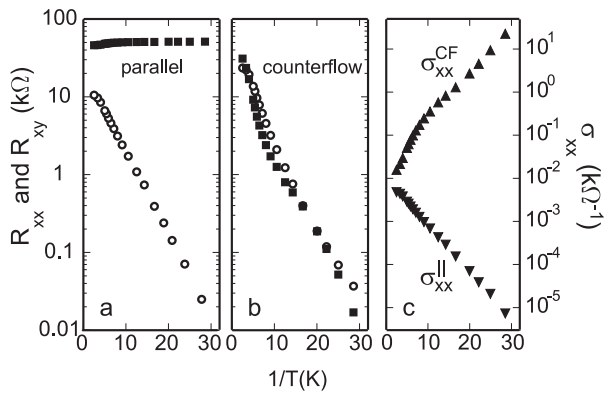}
\caption{\label{fig:fig4} Temperature dependences of various resistances and conductivities at \nnu\ and $d/\ell = 1.48$. a) Parallel current flow.  Open dots \Rpar, closed squares \RHpar. b) Counterflow.  Open dots \RCF, closed squares \RHCF. c) Parallel and counterflow longitudinal conductivities, \Spar\ and \SCF, respectively.}
\end{figure}

Figure 4c shows the parallel and counterflow conductivities, \Spar\ and \SCF, computed by inverting the resistivity tensor constructed from the data in Figs. 4a and 4b\cite{squares}. The computed parallel longitudinal conductivity \Spar\ falls with decreasing temperature, mirroring the behavior of \Rpar.  In contrast, \SCF\ {\it increases} steadily as the temperature falls.  At $T$ = 35 mK \SCF\ exceeds \Spar\ by more than 6 orders of magnitude.  In fact, \SCF\ at these low temperatures substantially exceeds both $e^2/h$ and the conductivity of the 2DES at zero magnetic field. 

These results support the prediction that the \nnu\ bilayer QHE state is a new kind of superfluid. Our data strongly suggest that counterflow electrical transport is dissipationless in the $T \rightarrow 0$ limit and that the longitudinal conductivity becomes very large in the same limit. However, our findings do not completely agree with the prevailing theory of counterflow transport at \nnu.  In the zero tunneling limit theory predicts that the \nnu\ state undergoes a Kosterlitz-Thouless transition at $T = T_{KT}$.  For $T < T_{KT}$ counterflow transport is expected to exhibit a strongly non-linear current dependence and, in the limit of zero current, be dissipationless.  Although $T_{KT}$ is estimated to be roughly 0.5K, and the onset temperature of the strong tunneling, Coulomb drag, and now counterflow anomalies at \nnu\ are consistent with this, we find no evidence of non-linearity of the observed counterflow longitudinal and Hall voltages at any temperature or excitation current thus far examined.  The origin of this discrepancy with theory is unknown, but we can speculate on some possibilities.

From a theoretical perspective the observation of finite dissipation below the KT transition suggests that unpinned free vortices are present in the disordered 2DES.  Such a possibility has already been raised by Fertig in connection with the finite interlayer tunneling conductance \cite{fertig}. Sheng, Balents and Wang have predicted that a new gauge-glass phase may exist in disordered bilayer 2DES systems at \nnu\ \cite{sheng}.  This new phase is expected to exhibit zero counterflow resistances, but only in the $T \rightarrow 0$ limit. It will be interesting to see how improvements in sample quality affect the experimentally observed dissipation in counterflow.

The experimental counterflow setup we employ ensures only that the {\it total} current flowing in the two layers are equal and oppositely directed.  It does not eliminate the possibility of local regions where the two currents are not precisely equal. If this occurs there will be a local parallel component of the transport which will be dissipative. Such a possibility might result from the inevitable inhomogeneities in the electron densities or a global effect associated with the geometry of the sample. The Hall bar was designed to reduce such effects, but nonetheless the current pathways for the two layers are not identical, especially near the ends of the bar where current is injected and withdrawn. Whether the excitonic superfluid would ``short-out'' regions of imperfect counterflow is an interesting question.

In conclusion, we have presented strong evidence that the strongly-coupled \nnu\ QHE phase in bilayer 2D electron systems is a counterflow superfluid.  The most dramatic aspect of our results is the vanishing of the Hall resistance at \nnu\ when oppositely directed currents flow in the two layers.  This observation is consistent with the theory of excitonic (or pseudospin) superfluidity at \nnu.  The origin of the excess dissipation we observe, which appears to vanish only as $T \rightarrow 0$, is an important unresolved question.

We thank A.H. MacDonald, S.M. Girvin, Z. Wang, and D.N. Sheng for enlightening conversations. This work was supported by the NSF under Grant No. DMR-0242946 and the DOE under Grant No. DE-FG03-99ER45766.


\begin{thebibliography}{99}
\bibitem{Hall} E.H. Hall, Am. J. Math. {\bf 2}, 287 (1879).

\bibitem{klitzing} K. von Klitzing, G. Dorda and M. Pepper, Phys. Rev. Lett. {\bf 45}, 494 (1980).

\bibitem{tsui} D.C. Tsui, H.L. Stormer, and A.C. Gossard, Phys. Rev. Lett. {\bf 48}, 1559 (1982).

\bibitem{supercon} Provided the field-induced vortices in the superconducting order parameter are pinned. 

\bibitem{perspectives} For a review, see the chapters by S.M. Girvin and A.H. MacDonald and by J.P. Eisenstein in {\it Perspectives in Quantum Hall Effects}, edited by A. Pinczuk and S. Das Sarma (Wiley, New York, 1997).

\bibitem{murphy} S.Q. Murphy, {\it et al.}, Phys. Rev. Lett. {\bf 72}, 728 (1994).

\bibitem{spielman1} I.B. Spielman, J.P. Eisenstein, L.N. Pfeiffer, and K.W. West, Phys. Rev. Lett. {\bf 84}, 5808 (2000) and Phys. Rev. Lett. {\bf 87}, 036803 (2001).

\bibitem{kellogg1} M. Kellogg, {\it et al.}, Phys. Rev. Lett. {\bf 88}, 126804 
(2002).

\bibitem{macd1} A.H. MacDonald, Physica (Amsterdam) {\bf 298B}, 129 (2001).

\bibitem{yang} Kun Yang, {\it et al.}, Phys. Rev. Lett. {\bf 72}, 732 (1994).

\bibitem{wenzee} X.G. Wen and A. Zee, Phys. Rev. Lett. {\bf 69}, 1811 (1992).

\bibitem{moon} K. Moon, {\it et al.}, Phys. Rev. B {\bf 51}, 5138 (1995).

\bibitem{ezawa}Z.F. Ezawa and A. Iwazaki, Phys. Rev. B{\bf 47}, 7295 (1993).

\bibitem{stern}A. Stern, S. Das Sarma, M.P.A. Fisher, and S.M. Girvin, Phys. Rev. Lett. {\bf 84}, 139 (2000).

\bibitem{jpe1} J.P. Eisenstein, L.N. Pfeiffer and K.W. West, Appl. Phys. Lett. {\bf 57}, 2324 (1990).

\bibitem{kellogg2} M. Kellogg, {\it et al.}, Phys. Rev. Lett. {\bf 90}, 246801 (2003).

\bibitem{squares} This computation assumes that the resistances \Rpar\ and \RCF\ are equal to the resistivities \rhopar\ and \rhoCF.  This is reasonable since the distance between the longitudinal voltage probes equals the width of the Hall bar but is substantially less than its length.

\bibitem{fertig} H.A. Fertig and J.P. Straley, Phys. Rev. Lett. {\bf 91}, 046806 (2003).

\bibitem{sheng} D.N. Sheng, L. Balents, and Ziqiang Wang, Phys. Rev. Lett. {\bf 91}, 116802 (2003).

\end{thebibliography}
\end{document}